\RequirePackage[l2tabu, orthodox]{nag}
\RequirePackage{ifpdf}
\RequirePackage[usenames,dvipsnames]{xcolor}
\RequirePackage{tikz}

\documentclass[prl,twocolumn,preprintnumbers,footinbib]{revtex4-1}%

\usepackage{hyperref}
\usepackage[utf8]{inputenc}%
\usepackage{mathtools, amscd, dsfont, amsthm, amssymb,array}
\usepackage{graphicx}
\usepackage[toc,page]{appendix}
\usepackage{etoolbox}
\mathtoolsset{showonlyrefs=false}
\usetikzlibrary{arrows,calc,shapes.misc,decorations.markings,snakes}
\tikzset{cross/.style={cross out, draw=black, minimum size=2*(#1-\pgflinewidth), inner sep=0pt,
        outer sep=0pt},cross/.default={3pt},
    gluon/.style={decorate, decoration={coil,aspect=0.9,segment length=5pt, amplitude=3pt}}}

\usepackage[english]{babel}
\usepackage{epsf,epsfig}
\usepackage{amssymb,amsmath,amsfonts}
\usepackage{mathrsfs}
\usepackage{color}
\usepackage{enumitem}

\usepackage{tikz}
\usepackage{bm}

\usepackage[mathscr]{eucal}

\def \be  {\begin{equation}}
\def \ee  {\end{equation}}
\def \ba  {\begin{eqnarray}}
\def \ea  {\end{eqnarray}}

\def \Tr {\mathop{\rm Tr}\nolimits}

\begin{document}

\title{BCJ relations in ${AdS}_5 \times S^3$ and the double-trace spectrum of super gluons}

\author{J.~M.~Drummond, R.~Glew, M.~Santagata}
\affiliation{
$\rule{0pt}{.01cm}$
\makebox[\textwidth][c]{School of Physics and Astronomy, University of Southampton, Highfield, SO17 1BJ, United Kingdom}\\
}

\begin{abstract}
\noindent
We revisit the four-point function of super gluons in $AdS_5 \times S^3$ in the spirit of the large $p$ formalism and show how the integrand of a generalised Mellin transform satisfies various non-trivial properties such as $U(1)$ decoupling identity, BCJ relations and colour-kinematic duality, in a way that directly mirrors the analogous relations in flat space. We unmix the spectrum of double-trace operators at large $N$ and find all anomalous dimensions at leading order. The anomalous dimensions follow a very simple pattern, resembling those of other theories with hidden conformal symmetries. 
\end{abstract}
\maketitle



\subsection{Introduction}
Understanding properties of (quantum) gravity theories and their relation to gauge theories is a primary goal in modern physics. Most of these relations are obscured in a lagrangian formulation, and seem to manifest all their majesty only through observables such as scattering amplitudes. 
The study of scattering amplitudes has led to a series of impressive and deep results, for example BCJ dualities \cite{Bern:2008qj} and double-copy constructions \cite{Bern:2010ue} in various theories, both at tree and loop-level (for a recent review, see \cite{Bern:2019prr}). The upshot is that there seems to be an underlying common structure between gravity and gauge theories, yet to be fully understood.
While most of these efforts have related to flat space, mainly because of the difficulties in performing such computations in curved backgrounds, a recent series of papers have begun the exploration of these properties in $AdS$ backgrounds, both in Mellin space \cite{Alday:2021odx,Zhou:2021gnu,Alday:2021ajh,Alday:2022lkk}, as well as in position \cite{Herderschee:2022ntr,Cheung:2022pdk} and momentum space \cite{Albayrak:2020fyp,Armstrong:2020woi}. Many of these developments have use bootstrap methods that have been highly successful in studying supergravity in $AdS$ \cite{Rastelli:2016nze,Rastelli:2017udc,Alday:2017xua,Aprile:2017xsp,Alday:2017vkk,Aprile:2017qoy,Caron-Huot:2018kta,Aprile:2018efk,Goncalves:2019znr,Aprile:2019rep,Alday:2019nin,Bissi:2020wtv,Aprile:2017bgs,Alday:2020dtb,Alday:2020lbp}, as well as string corrections \cite{Goncalves:2014ffa,Alday:2018kkw,Alday:2018pdi,Drummond:2019hel,Drummond:2020uni,Drummond:2019odu,Aprile:2020luw,Binder:2019jwn,Chester:2019pvm,Chester:2019,Chester:2020vyz,Drummond:2020dwr,Aprile:2020mus}.
In particular, in \cite{Zhou:2021gnu}, $AdS$ versions of colour-kinematic and double copy relations have been found. 

In this letter we further explore these relations by focusing on the four-point function of half-BPS operators dual to the scattering of four super gluons in $AdS_5 \times S^3$, first computed in \cite{Alday:2021odx}. In common with \cite{Zhou:2021gnu}, we will focus on the `reduced' Mellin amplitude (which manifests the supersymmetry of the theory) and cast this in a form which makes colour-kinematics and BCJ relations manifest by mirroring directly the form of the flat space amplitude. At leading order, the theory enjoys a hidden $8d$ conformal symmetry that nicely repackages all Kaluza-Klein modes into a simple reduced Mellin amplitude $\mathcal{M}_{\vec{p}}$. In notation inspired by the `large p' limit of \cite{Aprile:2020luw} it takes the form
\begin{equation}
\mathcal{M}_{\vec{p}} = \frac{n_s c_s}{{\bf s}+1}+\frac{n_t c_t}{{\bf t}+1}+\frac{n_u c_u}{{\bf u}+1}\,,
\end{equation}
where the kinematic ($n$) and colour ($c$) numerators obey the same Jacobi type relations.
We will see that the associated colour-ordered amplitudes also satisfy BCJ relations for \emph{all} Kaluza-Klein modes. These take the form,
\begin{align}
& ({{\bf t}+1}) \mathcal{M}_{\vec{p}}(1,2,3,4)=({{\bf u}+1}) \mathcal{M}_{\vec{p}}(1,3,4,2), \notag\\
& ({{\bf s}+1}) \mathcal{M}_{\vec{p}}(1,2,3,4)=({{\bf u}+1}) \mathcal{M}_{\vec{p}}(1,4,2,3), \notag \\
& ({{\bf t}+1}) \mathcal{M}_{\vec{p}}(1,4,2,3)=({{\bf s}+1})  \mathcal{M}_{\vec{p}}(1,3,4,2),
\end{align}
where $\mathcal{M}_{\vec{p}}(1,2,3,4)$ are the colour-ordered amplitudes.
The fact that these relations directly mirror their flat space counterparts is related to the existence of the $8d$ conformal symmetry. 

In the second part of the letter, we unmix the spectrum of double-trace operators exchanged in the OPE and compute all the anomalous dimensions at leading order. These CFT data are an important part of the bootstrap program for computing one-loop correlators beyond the lowest KK mode correlator. As shown in e.g. \cite{Alday:2017xua,Aprile:2017bgs,Aprile:2019rep,Alday:2021ajh}, the CFT data of the double-trace operators can be used to build the leading discontinuities of the correlator and then from them one can construct the full amplitude with the help of crossing symmetry.

In common with many other cases where the $AdS$ theory exhibits a hidden conformal symmetry, we find that the anomalous dimensions are given by a strikingly simple formula whose form we sketch here,
\begin{equation}
\eta_{\vec{\tau}}^{\pm}= - \frac{2}{N} \frac{\delta_{h,j}^{(2)}\delta_{\bar{h},j}^{(2)}}{(l_{8d}^{\pm}+1)_4}.
\end{equation}
Here the `effective' spin $l_{8d}$ and the quantity $\delta_{h,j}^{(2)}$ are functions of twist $\tau$, spin $l$ and $SU(2)\times SU(2)$ labels $[ab]$  of the double-trace operators and we will give their specific form later on.  The results are reminiscent of previous computations in other backgrounds \cite{Aprile:2021mvq,Aprile:2018efk,Abl:2021mxo}, and suggest that the hidden conformal symmetry, unavoidably, plays a primary role in constraining the data of these SCFTs.

\subsection{$AdS_5 \times S^3$ Mellin transform and the large $p$ formalism}
The $AdS_5 \times S^3$ background arises in two basic stringy setups. One can either consider a stack of $N$ D3-branes probing F-theory 7-brane singularities or a stack of $N_F$ D7-branes wrapping an $AdS_5 \times S^3$ subspace in the $AdS_5 \times S^5$ geometry of a stack of $N$ D3-branes. In both cases, the system preserves 8 supercharges, therefore the dual CFT is a $4d$ $\mathcal{N}=2$ theory with flavour group $G_F$, which we will keep generic because it is mostly irrelevant for the details considered in this paper. The low-energy degrees of freedom are those of a $\mathcal{N}=1$ vector multiplet which transforms in the adjoint of $G_F$. Upon reducing on the sphere, it provides an infinite tower of Kaluza-Klein modes organised in different multiplets. In the dual CFT, the super primaries of these multiplets are
half-BPS scalar operators of the form $\mathcal{O}_p^{I a_1 a_2 \ldots a_p; \bar{a}_1\bar{a}_2 \ldots \bar{a}_{p-2}}$.
Here $I$ is the colour index, $p$ is the scaling dimension of the operator, $a_1,\ldots, a_p$ are symmetrised $SU(2)_R$ R-symmetry indices and similarly $\bar{a}_i$ are indices of an additional $SU(2)_L$ flavour group; these last two groups realise the isometry group of the sphere $S^3$. In practice, and as usual in these contexts, is convenient to contract the indices with auxiliary bosonic two-component vectors $\eta$ and $\bar{\eta}$ to keep track of the $SU(2)_R \times SU(2)_L$ indices:
\begin{equation}
\mathcal{O}_p^{I} \equiv \mathcal{O}_p^{I;a_1 a_2 \ldots a_p; \bar{a}_1\bar{a}_2 \ldots \bar{a}_{p-2}}\eta_{a_1} \ldots \eta_{a_p} \bar{\eta}_{\bar{a}_1} \ldots \bar{\eta}_{\bar{a}_{p-2}}\,.
\end{equation}
In this paper we consider the amplitude of four super gluons, which we denote by
\begin{equation}
G_{\vec{p}}^{I_1 I_2 I_3 I_4}(x_i,\eta_i, \bar{\eta}_i)\equiv \langle \mathcal{O}_{p_1}^{I_1} \mathcal{O}_{p_2}^{I_2} \mathcal{O}_{p_3}^{I_3} \mathcal{O}_{p_4}^{I_4} \rangle.
\end{equation} 
A crucial point is that, in these theories, the strength of the self-gluon coupling is larger than the coupling of gluons to gravitons \cite{Alday:2021odx}. In light of this, one can perform an expansion in $1/N$ in which gravity is $1/N$ suppressed. Schematically, we have
\begin{equation}
G^{I_1 I_2 I_3 I_4}_{\vec{p}} =G_{\text{disc},\vec{p}}^{I_1 I_2 I_3 I_4}+\frac{1}{N} G_{\text{tree-gluon},\vec{p}}^{I_1 I_2 I_3 I_4} +\cdots
\end{equation} 
The first `disconnected' term is a sum over products of two-point functions and takes the form of (generalised) free theory. In terms of OPE data it contains the leading order contributions to the three-point functions of the external operators with exchanged two-particle operators. We will refer to the second term as the `tree-level' amplitude.

The correlator is subject to constraints due to superconformal symmetry. In particular, the superconformal Ward identites \cite{Nirschl:2004pa} allow us to split it into two parts, each separately respecting crossing symmetry,
\be
G_{\text{tree-gluon},{\vec{p}}} = G_{0,\vec{p}}  + \mathcal{P}\, \mathcal{I} \, \mathcal{A}_{\vec{p}}\,.
\ee
The term $G_{0,\vec{p}}$ contains all contributions due to protected multiplets at this order in $1/N$. The second term contains all the logarithmic terms which arise due to two-particle operators receiving anomalous dimensions. 
It contains certain kinematic factors $\mathcal{P}$ and $\mathcal{I}$, due to bosonic and femionic symmetries respectively. First, let us define the propagator via 
\be
g_{ij}=\frac{y_{ij}^2}{x_{ij}^2}\, 
\ee
where $y_{ij}^2 = \langle \eta_i \eta_j\rangle \langle \bar{\eta}_i \bar{\eta}_j\rangle$ with $\langle \eta_i \eta_j\rangle = \eta_{ia} \eta_{jb} \epsilon^{ab}$ and similarly $\langle \bar{\eta}_i \bar{\eta}_j\rangle = \bar{\eta}_{i\bar{a}} \bar{\eta}_{j\bar{b}} \epsilon^{\bar{a}\bar{b}}$.
We also introduce cross-ratios via
\begin{align}
\frac{x_{12}^2 x_{34}^2}{x_{13}^2 x_{24}^2} &= U =x \bar{x}\,, &&\frac{x_{14}^2 x_{23}^2}{x_{13}^2 x_{24}^2} = V =(1-x)(1-\bar{x})\,, \notag \\ 
\frac{y_{12}^2 y_{34}^2}{y_{13}^2 y_{24}^2} &= \tilde{U}=y \bar{y}\,, &&\frac{y_{14}^2 y_{23}^2}{y_{13}^2 y_{24}^2} = \tilde{V}=(1-y)(1-\bar{y})\,. \notag
\end{align}
Note that we can write the $y,\bar{y}$ variables in terms of the $\eta$ and $\bar{\eta}$ variables as
\be
y = \frac{\langle \eta_1 \eta_2 \rangle \langle \eta_3 \eta_4 \rangle}{\langle \eta_1 \eta_3 \rangle \langle \eta_2 \eta_4 \rangle}\, \qquad \bar{y} = \frac{\langle \bar{\eta}_1 \bar{\eta}_2 \rangle \langle \bar{\eta}_3 \bar{\eta}_4 \rangle}{\langle \bar{\eta}_1 \bar{\eta}_3 \rangle \langle \bar{\eta}_2 \bar{\eta}_4 \rangle}\,.
\ee
The kinematic factors are then given by
\begin{align}
\mathcal{P} \equiv \frac{g_{12}^{k_s} g_{14}^{k_t}  g_{24}^{k_u}  \big( g_{13} g_{24} \big)^{p_3} }{\langle \bar{\eta}_1 \bar{\eta}_3\rangle^2 \langle \bar{\eta}_2 \bar{\eta}_4\rangle^2}\,, \quad \mathcal{I} = (x - y)(\bar{x}-y)\,,
\end{align}
where
\begin{align}
k_s \!=\! \tfrac{ p_1+p_2-p_3-p_4}{2},\,\, k_t\!=\! \tfrac{p_1+p_4-p_2-p_3}{2},\,\, k_u \!=\! \tfrac{ p_2+p_4-p_3-p_1}{2}.\notag
\end{align}
Note that, due to the presence of the factor $\mathcal{I}=(x- y)(\bar{x}-y)$, the remaining function $\mathcal{A}_{\vec{p}}^{I_1 I_2 I_3 I_4}$ has the same degree in $y, \bar{y}$. Moreover, since $\mathcal{A}_{\vec{p}}^{I_1 I_2 I_3 I_4}$ is symmetric under $y, \bar{y}$ exchange, we can write it as a function of $\tilde U,\tilde V$ as well as $U$ and $V$ and the charges $\vec{p}$.

The function $\mathcal{A}_{\vec{p}}^{I_1 I_2 I_3 I_4}$ admits a very compact and natural representation, that extends the well known Mellin transform \cite{Mack:2009mi,Penedones:2010ue} to the compact space. The transform makes manifest the so-called large $p$ limit \cite{Aprile:2020luw} - where $p$ here refers to the charges - and it was found to be very useful in the context of $AdS_5 \times S^5$ \cite{Aprile:2020luw,Aprile:2020mus} and $AdS_3 \times S^3$ backgrounds \cite{Aprile:2021mvq}. In our conventions the generalised Mellin transform $\mathcal{M}$ is defined via 
\begin{equation}\label{app_amplitude_def}
\mathcal{A}_{\vec{p}}^{I_1 I_2 I_3 I_4} \!= -\oint\!ds dt \!\oint\!d\tilde{s} d\tilde{t} \,U^s V^t \tilde{U}^{\tilde s} \tilde{V}^{\tilde t} \,\Gamma\, \mathcal{M}_{\vec{p}}^{I_1 I_2 I_3 I_4} 
\end{equation}
where $\mathcal{M}_{\vec{p}}^{I_1 I_2 I_3 I_4} \equiv \mathcal{M}_{\vec{p}}^{I_1 I_2 I_3 I_4} (s,t,\tilde s, \tilde t)$.
The kernel $\Gamma$ is factorised into $AdS_5$ and $S^3$ contributions and takes the form $\Gamma = \mathfrak{S} \,\Gamma_s \Gamma_t \Gamma_u$ with
\begin{align}
\mathfrak{S}=\pi^2 \tfrac{\ (-)^{\tilde t}(-)^{\tilde u} }{\sin(\pi \tilde t\, )\sin(\pi \tilde u)}\,, \quad \Gamma_s = \tfrac{ \Gamma[-s]\Gamma[-s+k_s]}{\Gamma[1+\tilde s]\Gamma[1+\tilde s + k_s]}
\end{align}
and $\Gamma_t$, $\Gamma_u$ defined similarly.
Note that the Mellin variables obey the  relations,
\ba \label{onsh}
s+t+u=-p_3-1, \quad \tilde s+\tilde t +\tilde u = p_3-2\,,
\ea
which may be used to eliminate $u$ and $\tilde{u}$.
Note also that the amplitude $\mathcal{A}_{\vec{p}}^{I_1 I_2 I_3 I_4}$ is polynomial in $\tilde U$ and $\tilde V$. In fact, the integral over $\tilde{s}$, $\tilde{t}$ can be turned into a discrete sum over a certain domain that in our case is given by
\be
\label{triangle}
\textit{T}=\{  \tilde s\ge max(0,-k_s),\  \tilde t,\tilde u\ge 0 \}\,.
\ee
The contour integral in $s$ and $t$ requires a little care and we will return to this point in the next section. The double integral (\ref{app_amplitude_def}), when combined with the amplitude $\mathcal{M}_{\vec{p}}$ given in the next section, precisely coincides with the result given in \cite{Alday:2021odx}.

This generalised $AdS_5 \times S^3$ Mellin transform is quite useful because, as shown in \cite{Aprile:2020luw}, in the large $p$ limit, the integrals localise on a classical saddle point. The authors show that the computation matches with that of four geodesics shooting from the boundary and meeting in a common bulk point at which the particles scatter as if they were in flat space. At the saddle point, the `bold-face' variables
\begin{equation}
{\bf s}=s+\tilde{s}, \quad {\bf t}=t+\tilde{t}, \quad {\bf u}=u+\tilde{u}, \quad {\bf s}+{\bf t}+{\bf u}=-3
\end{equation}
become proportional to the flat space Mandelstam variables. This explains why, for large $p$,  the Mellin amplitude $\mathcal{M}_{\vec{p}}^{I_1 I_2 I_3 I_4}$ is fixed by the flat space S-matrix with the Mandelstam variables replaced by the bold-face variables ${\bf s},{\bf t},{\bf u}$.

Moreover, as we will see, the integrand $\mathcal{M}_{\vec{p}}^{I_1 I_2 I_3 I_4}$ satisfies BCJ and double-copy relations, directly analogous to the flat space relations, incorporating all Kaluza-Klein modes.

\subsection{BCJ and colour-kinematics in $AdS_5 \times S^3$}
Let us consider the field theory amplitude computed in \cite{Alday:2021odx} within this formalism. As in \cite{Zhou:2021gnu} we consider the reduced Mellin amplitude $\mathcal{M}_{\vec{p}}$. In the colour-factor basis, the amplitude $\mathcal{M}_{\vec{p}}$ takes the following very simple form when written in terms of the bold-face variables,
\begin{equation}\label{supergluonsA}
\mathcal{M}_{\vec{p}}^{I_1 I_2 I_3 I_4} = \frac{n_s c_s}{{\bf s}+1}+\frac{n_t c_t}{{\bf t}+1}+\frac{n_u c_u}{{\bf u}+1}\,.
\end{equation}
Here we have
\begin{align}
& n_s = \frac{1}{3}\left( \frac{1}{{\bf t}+1}- \frac{1}{{\bf u}+1} \right), \qquad c_s = f^{I_1 I_2 J}f^{I_3 I_4 J}, \notag \\
& n_t = \frac{1}{3} \left( \frac{1}{{\bf u}+1}- \frac{1}{{\bf s}+1}\right), \qquad c_t = f^{I_1 I_4 J}f^{I_2 I_3 J}, \notag \\
& n_u =\frac{1}{3} \left(\frac{1}{{\bf s}+1}- \frac{1}{{\bf t}+1}\right), \qquad c_u =f^{I_1 I_3 J}f^{I_2 I_4 J}.
\end{align}

As described above, the large $p$ limit ensures that the amplitude reduces to the flat amplitude with the Mandelstam replaced by bold-face variables
\begin{equation}
\begin{split}
& \mathcal{M}_{\vec{p}}^{I_1 I_2 I_3 I_4}\xrightarrow[s,t,\tilde{s},\tilde{t},\vec{p} \rightarrow \infty]{} \mathcal{V}_{\text{YM}}^{I_1 I_2 I_3 I_4}({\bf s} , {\bf t}),  \\
& \mathcal{V}_{\text{YM}}^{I_1 I_2 I_3 I_4}({\bf s} , {\bf t})=  \frac{n_s c_s}{{\bf s}}+\frac{n_t c_t}{{\bf t}}+\frac{n_u c_u}{{\bf u}}, \\
&  n_s = \frac{1}{3} \left(\frac{1}{{\bf t}}- \frac{1}{{\bf u}} \right),\ldots
\end{split}
\end{equation}
where $\mathcal{V}_{\text{YM}}^{I_1 I_2 I_3 I_4}$ is the field theory gluon amplitude in flat space, see e.g. \cite{Elvang:2015rqa}.
Note that this limit somewhat restores the symmetry between $AdS$ and $S$; in this sense it is a generalisation of the usual flat space limit in which only the $AdS$ (Mellin) variables $s,t$ are taken to be large.

In principle, away from large $p$, nothing would prevent the amplitude to depend on $s, \tilde{s}, \cdots$ separately. However, from (\ref{supergluonsA}) we see that in fact the full amplitude $\mathcal{M}$ is just a function of the bold face variables. This fact is a consequence of a hidden $8d$ conformal symmetry of the amplitude. This symmetry allows one to promote the correlator $\mathcal{M}_{2222}^{I_1 I_2 I_3 I_4}$ to a generating function for correlators with arbitrary charges $\vec{p}$. Then, $\mathcal{M}_{\vec{p}}^{I_1 I_2 I_3 I_4}$ follows from `covariantising' $\mathcal{M}_{2222}^{I_1 I_2 I_3 I_4}$:
\begin{equation}
\mathcal{M}_{2222}^{I_1 I_2 I_3 I_4}(s,t) \xrightarrow{\texttt{8d-symm}}  \mathcal{M}_{\vec{p}}^{I_1 I_2 I_3 I_4}= \mathcal{M}_{2222}^{I_1 I_2 I_3 I_4}({\bf s}, {\bf t}). \notag
\end{equation}
These features are entirely analogous to $AdS_3 \times S^3$ \cite{Rastelli:2019gtj,Giusto:2019pxc,Giusto:2020neo,Aprile:2021mvq} and $AdS_5 \times S^5$ \cite{Caron-Huot:2018kta,Aprile:2020luw} backgrounds where the dynamics is also controlled by hidden conformal symmetries. In other words, $ \mathcal{A}_{\vec{p}}^{I_1 I_2 I_3 I_4}$  is generated from $ \mathcal{A}_{2222}^{I_1 I_2 I_3 I_4}$ upon acting with a differential operator which takes a very simple form. In fact, we can give a general formula of the operator that interpolates between the three cases. Parametrising the space as $AdS_{\theta_1+1} \times S^{\theta_2+1}$, the amplitude $\mathcal{A}$ for general $\vec{p}$ is generated from the one with the lowest charges $\vec{p} = (qqqq)$ with $q=\tfrac{\theta_1}{2}$ via a differential operator,
\begin{equation}
 \mathcal{A}_{\vec{p}}^{I_1 I_2 I_3 I_4}= \mathcal{D}_{\vec{p}}^{\theta_1,\theta_2} \Bigl [ U^{\tfrac{\theta_1+\theta_2}{2}} \mathcal{A}_{qqqq}^{I_1 I_2 I_3 I_4} \Bigr]\,, \quad q=\tfrac{\theta_1}{2}\,.
\end{equation}
The operator $\mathcal{D}_{\vec{p}}^{\theta_1,\theta_2}$ takes the following form
\begin{equation}
\mathcal{D}_{\vec{p}}^{\theta_1,\theta_2}= U^{-\tfrac{\theta_1+\theta_2}{2}} \sum_{\tilde{s},\tilde{t}} \biggl(\frac{\tilde{U}}{U}\biggr)^{\tilde{s}}\biggl(\frac{\tilde{V}}{V} \biggr)^{\tilde{t}} \hat{\mathcal{D}}_{\vec{p},\tilde{s},\tilde{t}}^{\theta_1,\theta_2} 
\end{equation}
where 
\begin{align}
\label{operatorD}
\hat{\mathcal{D}}_{\vec{p},\tilde{s},\tilde{t}}^{\theta_1,\theta_2} \! = \!\! \prod_{a=\{0, k_s\}} \!\! & { \tfrac{ (U\partial_U +1-\tfrac{\theta_1+\theta_2}{2} - \tilde s -a)_{\tilde s+a}}{(-)^a (\tilde s+a)!}}   \\
								 \times   \! \prod_{ b=\{0, k_t\} }   \!\! &{ \tfrac{ (V\partial_V +1- \tilde t -b)_{\tilde t+b}}{(-)^b (\tilde t+b)!}}
								  \prod_{c=\{0, k_u\}}  \!\! {\tfrac{ (U\partial_U+V\partial_V)_{\tilde u+c}}{ (\tilde u+c)!}} \notag
\end{align}
%
and we turned the sphere integral into a sum restricted to the domain $T$. The operator transforms the gamma functions of $\mathcal{A}_{qqqq}^{I_1 I_2 I_3 I_4}$ into those of $\mathcal{A}_{\vec{p}}^{I_1 I_2 I_3 I_4}$  and replaces $s,t,u$ with ${\bf s},{\bf t},{\bf u}$, as it can be easily checked using (a consequence of) Euler's reflection identity.

In fact, as observed above, with the $AdS_5 \times S^3$ background, our variables obey ${\bf s} + {\bf t} + {\bf u} = -3$. Therefore the the Mellin amplitude $\mathcal{M}$ is literally the same function as the flat space amplitude with the Mandelstam variables $s,t,u$ replaced by the shifted bold face variables $({\bf s} +1)$, $({\bf t}+1)$, $({\bf u}+1)$. It follows immediately that all the relations obeyed by the flat space amplitudes also apply to $\mathcal{M}$. Note that it is not trivial that this holds; for example, the analogous relation for $AdS_5 \times S^5$ is ${\bf s} + {\bf t} + {\bf u} = -4$ \cite{Aprile:2020luw}.
As an example of the properties obeyed by $\mathcal{M}$ we have that
\begin{align}
n_s + n_t + n_u & =0\,, \notag \\
c_s + c_t + c_u & =0\,,
\label{CKduality}
\end{align}
which gives an $AdS$ version of the colour-kinematic duality, which was already observed in \cite{Zhou:2021gnu}. Note that (\ref{CKduality}) captures this duality for all Kaluza-Klein modes.
This duality is intimately connected with the so-called BCJ relations between colour-ordered amplitudes. 
Recall that the full colour-dressed amplitude is:
\begin{equation}\label{colourorderedamp}
\mathcal{M}_{\vec{p}}^{I_1 I_2 I_3 I_4}=\sum_{\mathcal{P}(2,3,4)} \Tr \left( T^{I_1}T^{I_2}T^{I_3}T^{I_4} \right)\mathcal{M}_{\vec{p}}(1,2,3,4)
\end{equation}
where the partial amplitudes $\mathcal{M}_{\vec{p}}(1,2,3,4)$ are the colour-ordered amplitudes and ${\mathcal{P}(2,3,4)}$ are the permutations of points $(2,3,4)$.
The translation from one basis to another is:
\begin{align}
 c_s= & \Tr \left( T^{I_1}T^{I_2}T^{I_3}T^{I_4} \right)+\Tr \left( T^{I_1}T^{I_4}T^{I_3}T^{I_2} \right)  \notag \\
& -\Tr \left( T^{I_1}T^{I_2}T^{I_4}T^{I_3} \right)-\Tr \left( T^{I_1}T^{I_3}T^{I_4}T^{I_2} \right),  \notag\\
 c_t=& \Tr \left( T^{I_1}T^{I_4}T^{I_2}T^{I_3} \right)+\Tr \left( T^{I_1}T^{I_3}T^{I_2}T^{I_4} \right)- \notag\\
&\Tr \left( T^{I_1}T^{I_4}T^{I_3}T^{I_2} \right)-\Tr \left( T^{I_1}T^{I_2}T^{I_3}T^{I_4} \right), \notag\\
c_u = & \Tr \left( T^{I_1}T^{I_3}T^{I_4}T^{I_2} \right)+\Tr \left( T^{I_1}T^{I_2}T^{I_4}T^{I_3} \right)- \notag\\
& \Tr \left( T^{I_1}T^{I_3}T^{I_2}T^{I_4} \right)-\Tr \left( T^{I_1}T^{I_4}T^{I_2}T^{I_3} \right).
\end{align}
The colour-ordered amplitudes then read as follows,
\begin{align}
& \mathcal{M}_{\vec{p}} (1,2,3,4)= \mathcal{M}_{\vec{p}} (1,4,3,2)=\frac{n_s}{{\bf s}+1}-\frac{n_t}{{\bf t}+1} \,,\notag \\
& \mathcal{M}_{\vec{p}} (1,2,4,3)= \mathcal{M}_{\vec{p}} (1,3,4,2)= \frac{n_u}{{\bf u}+1}-\frac{n_s}{{\bf s}+1} \,,\notag \\
& \mathcal{M}_{\vec{p}} (1,3,2,4)= \mathcal{M}_{\vec{p}} (1,4,2,3)= \frac{n_t}{{\bf t}+1}-\frac{n_u}{{\bf u}+1}\, .
\label{colourord}
\end{align}
All the relations obeyed by the flat space colour-ordered amplitudes obviously also hold here. In particular, an $n$-point function satisfies cyclicity, reflection, Kleiss-Kuijf relations and the $U(1)$ decoupling identity that reduce the number of independent colour-ordered amplitudes to $(n-2)!$. These last two relations coincide for a four-point function.
From \eqref{colourord}, we can see that an analogous $U(1)$ decoupling identity holds here:
\begin{equation}\label{decouplingidentity}
\mathcal{M}_{\vec{p}}(1,2,3,4)+\mathcal{M}_{\vec{p}} (1,2,4,3) + \mathcal{M}_{\vec{p}} (1,3,2,4)=0.
\end{equation}
There are further relations, known as BCJ relations, that reduce the number of independent colour-ordered amplitudes to $(n-3)!$,
\begin{align}
& ({{\bf t}+1}) \mathcal{M}_{\vec{p}}(1,2,3,4)=({{\bf u}+1}) \mathcal{M}_{\vec{p}}(1,3,4,2) \,,\notag\\
& ({{\bf s}+1}) \mathcal{M}_{\vec{p}}(1,2,3,4)=({{\bf u}+1}) \mathcal{M}_{\vec{p}}(1,4,2,3) \,,\notag \\
& ({{\bf t}+1}) \mathcal{M}_{\vec{p}}(1,4,2,3)=({{\bf s}+1}) \, \mathcal{M}_{\vec{p}}(1,3,4,2)\,,
\label{BCJrels}
\end{align}
where we used the on-shell relation ${\bf s}+{\bf t}+{\bf u}=-3$. We stress again that the relations (\ref{BCJrels}) capture the appearance of BCJ relations in $AdS$ for \emph{all} Kaluza-Klein modes. Such relations are manifest at level of the reduced Mellin amplitude while they do not hold, at least directly, for the full Mellin amplitude \cite{Alday:2022lkk}. It is an interesting open question how such relations might extend to higher point amplitudes in $AdS$ and what the role of a reduced Mellin amplitude might be in this regard.

Having introduced the colour-ordered amplitudes, let us return to the issue of the contour in the Mellin integral (\ref{app_amplitude_def}). It should be noted that the presence of poles at ${\bf s}=-1$, ${\bf t}=-1$ and ${\bf u}=-1$ is potentially a problem for the contour of integration. In fact, since ${\bf s} + {\bf t} + {\bf u}=-3$, the simultaneous presence of these poles leaves no region in the real $s,t$ plane for the contour to pass through, while separating left moving and right moving sequences of poles in the Mellin integrand. Thus the same property which leads to the direct analogy with the flat space amplitudes also leads to a subtlety in returning to position space from Mellin space. For the colour ordered amplitudes, one does not have all three poles present simultaneously. Thus we propose that the correct definition for the contour is tied to the colour-ordering and we define analogously a colour-ordered correlator,
\begin{equation}
\mathcal{A}(1,2,3,4) \!= -\oint\!ds dt \!\oint\!d\tilde{s} d\tilde{t} \,U^s V^t \tilde{U}^{\tilde s} \tilde{V}^{\tilde t} \,\Gamma\, \mathcal{M}(1,2,3,4)\,, \notag
\end{equation}
The contour can now be taken to lie slightly below ${\bf s} = -1$ and ${\bf t} =-1$. Note then that this introduces a subtlety in interpreting the BCJ relations (\ref{BCJrels}) back in position space, since the left and right hand sides of these equations are to be integrated over slightly different contours.

To conclude, let us point out that there is also an $AdS$ version of the double-copy prescription \cite{Zhou:2021gnu}. Replacing colour with kinematic factors we get
\begin{align}
\mathcal{M}_{\vec{p}}^{I_1 I_2 I_3 I_4}  \xrightarrow[ c_i \rightarrow  n_i]{}& \frac{n_s^2}{{\bf s}+1}+\frac{n_t^2}{{\bf t}+1}+\frac{n_u^2}{{\bf u}+1} \\
= &\frac{1}{({\bf s}+1)({\bf t}+1)({\bf u}+1)} \propto \mathcal{M}_{\vec{p}}^{\text{SUGRA}}. \notag 
\end{align}
This is nothing but the SUGRA amplitude in $AdS_5 \times S^5$ \cite{Rastelli:2016nze} rewritten in the large $p$ formalism  \cite{Aprile:2020luw}, upon reinterpreting ${\bf s},{\bf t},{\bf u}$ as the $\mathcal{N}=4$ variables, i.e. subject to the constraint  ${\bf u}= -{\bf s}-{\bf t}-4$. 
Note also that, similarly to flat space \cite{Bern:2019prr}, we can use BCJ and colour-kinematic duality to derive an $AdS$ version of the KLT relations:
\begin{equation}
\mathcal{M}_{\vec{p}}^{\text{SUGRA}}= ({\bf s} +1) \mathcal{M}_{\vec{p}}(1,2,3,4) \mathcal{M}_{\vec{p}}(1,2,4,3).
\end{equation}

\subsection{Long disconnected free theory}
The rest of the letter will be devoted to investigate the structure of the anomalous dimensions of the double-trace operators exchanged in the OPE at large $N$. 
In order to do so, we need two ingredients: the superconformal block decomposition of disconnected generalised free theory and that of the $\log U$ discontinuity of the tree-level correlator. The anomalous dimensions are then nothing but the eigenvalues of a certain matrix built out of the block coefficients of these two decompositions.
On top of the above mentioned (usual) technology, we also have to deal with the non-trivial flavour structure of the amplitude. However, since all of this just amounts to considering certain symmetric or antisymmetric combinations built out of the correlator, we postpone the discussion on flavour structures to the end of next section. A more detailed discussion can be found in \cite{Alday:2021ajh}.

Let us begin with disconnected free theory. The only correlators with non-zero disconnected contributions are with pairwise equal charges and their spacetime dependence can be computed by performing simple Wick contractions. We have
\begin{align}\label{discocorr}
G_{\text{disc},pqpq}^{I_{1}I_{2}I_{3}I_{4}}= & \delta^{I_{1}I_{2}}\delta^{I_{3}I_{4}} \delta_{pq} \frac{g_{12}^{p} g_{34}^{p}}{\langle \bar{\eta}_1 \bar{\eta}_2\rangle^2 \langle \bar{\eta}_3 \bar{\eta}_4 \rangle^2} \notag \\
+ &\underbrace{\delta^{I_{1}I_{3}}\delta^{I_{2}I_{4}} \frac{g_{13}^{p} g_{24}^{p}}{\langle \bar{\eta}_1 \bar{\eta}_3\rangle^2 \langle \bar{\eta}_2 \bar{\eta}_4 \rangle^2}}_{\texttt{u-channel}} \notag \\
+&\underbrace{\delta^{I_{1}I_{4}}\delta^{I_{2}I_{3}} \delta_{pq} \frac{g_{14}^{p} g_{23}^{p}}{\langle \bar{\eta}_1 \bar{\eta}_4\rangle^2 \langle \bar{\eta}_2 \bar{\eta}_3 \rangle^2} }_{\texttt{t-channel}}.
\end{align}
However, due to the non-trivial colour structure of the amplitude, only representations with a definite parity under $t\leftrightarrow u$ exchange enter the OPE. In practice, we need to decompose the following combinations of diagrams
\begin{equation}\label{discocorr1}
G_{\text{disc},pqpq}^\pm= \delta_{pq}\frac{g_{14}^{p} g_{23}^{p}}{\langle \bar{\eta}_1 \bar{\eta}_4\rangle^2 \langle \bar{\eta}_2 \bar{\eta}_3 \rangle^2}
\pm \frac{g_{13}^{p} g_{24}^{p}}{\langle \bar{\eta}_1 \bar{\eta}_3\rangle^2 \langle \bar{\eta}_2 \bar{\eta}_4 \rangle^2}
\end{equation} 
Now, following \cite{Nirschl:2004pa}, we first extract the unprotected contribution and then decompose it in long superblocks \cite{Doobary:2015gia}, whose form is given in the appendix. 
The block decomposition reads
\begin{equation}\label{discoeq}
G_{\text{disc},pqpq}^{\pm} \big |_{\text{long}}= \sum_{\vec{\tau}}  {L}_{\vec{\tau}}^\pm \mathbb{L}_{\vec{\tau}},
\end{equation}
where $\mathbb{L}_{\vec{\tau}}$ are the long superblocks. 
We find that the coefficients take a particularly simple form,
\begin{equation}\label{longblockcoeff}
L_{\vec{\tau}}^{\pm} = -\frac{\pm 1+ (-1)^{a+l} \delta_{p q}}{(p-1)(q-1)}A_h A_{\bar{h}}B_j B_{\bar{j}} \boldsymbol\delta  \,.
\end{equation}
Here the $A$ and $B$ factors are given by 
\begin{align}
A_h= &\frac{\Gamma(h+\tfrac{p-q}{2})\Gamma(h-\tfrac{p-q}{2})\Gamma(h+\frac{p+q}{2}-1)}{\Gamma(2h-1)\Gamma(h-\frac{p+q}{2}+1)}, \\
 B_j= &\frac{\Gamma(2-2j)}{\Gamma(1-j+\tfrac{p-q}{2})\Gamma(1-j-\tfrac{p-q}{2})}  \notag\\
  & \times \frac{1}{\Gamma(\frac{p+q}{2}+j-1)\Gamma(\frac{p+q}{2}-j)} \,, \notag
\end{align}
while $\boldsymbol\delta$ is given by
\begin{equation}
 \boldsymbol\delta = \frac{ \delta_{h,j}^{(2)}-\delta_{\bar{h},j}^{(2)} }{\delta_{h,j}^{(2)}\delta_{\bar{h},j}^{(2)}}, \qquad \delta_{\bar{h},j}^{(2)}= (h-j)(h+j-1).
\end{equation}
Here, $h,\bar{h}$ and $j,\bar{j}$ label, respectively, the conformal and internal representations. We can also express them in terms of the more common quantum labels $\vec{\tau}=(\tau,b,l,a)$
\begin{equation}\label{hlabels}
h=\frac{\tau}{2}+1+l, \quad \bar{h}=\frac{\tau}{2}, \quad j= -\frac{b}{2}-a,   \quad \bar{j}= -\frac{b}{2},
\end{equation}
where $\tau,l$ are twist and spin, and $b,a$ can be seen as the analogues of twist and spin on the sphere.
Note the different ways the two internal $SU(2)$ factors enter the coefficients. On the one hand, $SU(2)_L$ only comes in through the function $B_{\bar{j}}$. On the other hand, the decomposition under the R-symmetry group $SU(2)_R$ produces also the function $\boldsymbol\delta$ and, in particular the combination ${\delta_{h,j}^{(2)}\delta_{\bar{h},j}^{(2)}}$.
This object is the eigenvalue of a Casimir operator operator acting on the blocks,
\begin{align}
\label{delta4op}
& \mathcal{D}_4 \bigl( U^{\tfrac{p_{43}}{2}}\tilde{U}^{2-\tfrac{p_{43}}{2}}(x- \bar{x})  \mathcal{G}_{\tau,l} \mathcal{H}_{b,a} \bigr) \notag \\
&=  \delta_{h,j}^{(2)}\delta_{\bar{h},j}^{(2)} \, \, \bigl( U^{\tfrac{p_{43}}{2}}\tilde{U}^{2-\tfrac{p_{43}}{2}}(x -\bar{x}) \mathcal{G}_{\tau,l} \mathcal{H}_{b,a} \bigr) \,.
\end{align}
Here the differential operator $\mathcal{D}_4$ is given by
\begin{equation*}
\mathcal{D}_4 \!=\! (D_{x}^+-D_{y}^-) (D_{\bar{x}}^+-D_{y}^-),\,\, D_{x}^\pm \mathcal{F}_h^\pm (x) \!=\! h(h-1) \mathcal{F}_h^\pm (x),
\end{equation*}
where $D_{x}^{\pm}$ is \cite{Dolan:2011dv},
\begin{equation}
D_{x}^{\pm}= x^2 \partial_x(1-x)\partial_x \pm (p_{12}+p_{34})x^2 \partial_x- p_{12}p_{34} \,x.
\end{equation}
and the functions $ \mathcal{G}_{\tau,l}, \, \mathcal{H}_{b,a},\,\mathcal{F}_h^\pm$, that appear in the long superblocks, are defined in the appendix. 
Note that $\mathcal{F}_{\bar{j}}^-(\bar{y})$ is a spectator in \eqref{delta4op}.
The presence of ${\delta_{h,j}^{(2)}\delta_{\bar{h},j}^{(2)}}$ suggests that the hidden symmetry in free theory is realised not on the correlator of the $\mathcal{O}_{p}$ but on a correlator of superconformal descendants of $\mathcal{O}_p$, obtained by action of the Casimir. A more detailed discussion can be found in \cite{Caron-Huot:2018kta} for $AdS_5 \times S^5$ and in \cite{Abl:2021mxo} for $AdS_2 \times S^2$ background, where the logic is exactly the same. In these last two cases, $\mathcal{D}_4$ is replaced by $\mathcal{D}_8$ and $\mathcal{D}_2$, respectively.

In the rest of the section we would like to highlight some features common to various $AdS_{\theta_1+1} \times S^{\theta_2+1}$ backgrounds.
To start with, the coefficients of long disconnected free theory are very similar in all these theories, (c.f. formulas in \cite{Aprile:2021mvq}). In fact, upon shifting $(p,q)\rightarrow (p+1, q+1)$ in \eqref{longblockcoeff}, they are the same when written in $h-$type variables, except for the function $\boldsymbol\delta$ which depends on the theory,
\begin{align}
&\frac{1}{{\boldsymbol \delta}}=\frac{\delta_{h,\bar{h},j,\bar{j}}^{(4)}\delta_{h,\bar{h},\bar{j},j}^{(4)}} {\delta_{h,\bar{h},j,\bar{j}}^{(4)}+\delta_{h,\bar{h},\bar{j},j}^{(4)}},  \qquad \qquad AdS_3 \times S^3, \notag\\
&\frac{1}{{\boldsymbol \delta}}=\frac{\delta_{h,\bar{h},j,\bar{j}}^{(4)}\delta_{h,\bar{h},\bar{j},j}^{(4)}} {\delta_{h,\bar{h},j,\bar{j}}^{(4)}-\delta_{h,\bar{h},\bar{j},j}^{(4)}},  \qquad \qquad AdS_5 \times S^5, \notag \\
& \frac{1}{{\boldsymbol \delta}} = \frac{\delta_{h,j}^{(2)}\delta_{\bar{h},j}^{(2)}}{\delta_{h,j}^{(2)}-\delta_{\bar{h},j}^{(2)}}, \qquad   \qquad \qquad AdS_5 \times S^3. 
\end{align}
where $\delta_{h,\bar{h},j,\bar{j}}^{(4)} \equiv \delta_{h,j}^{(2)} \delta_{h,\bar{j}}^{(2)}$.

Note also that the dictionary between $\vec{h}$ labels and $\vec{\tau}$ labels depends on the theory. By borrowing the results from \cite{Aprile:2021mvq}, we can write down the general dictionary interpolating between the three backgrounds
\begin{equation}
\begin{split}
& h=\frac{\tau+\theta_2}{2}+l,\quad \bar{h}=\frac{\tau+\theta_2-\theta_1}{2}+1, \\ 
& j= -\frac{b+\theta_2}{2}-a+1,   \quad \bar{j}= -\frac{b}{2}.
\end{split}
\end{equation}
The existence of such formulas for disconnected graphs interpolating between different theories turns out to be a particular case of a more general formula for all free-theory diagrams which can be proved through a Cauchy identity \cite{Aprile:2021pwd}.

\subsection{Anomalous dimensions and residual degeneracy}
We will not give too many details of the computation, which can be found in \cite{Aprile:2017xsp,Aprile:2018efk}  for the similar $AdS_5 \times S^5$ case; analogous computations in $AdS_3 \times S^3$ can be found in \cite{Aprile:2021mvq}. The main difference with the $\mathcal{N}=4$ case is that here double-trace operators have a flavour structure. Because of this, there will be two types of anomalous dimensions, those of operators exchanged in symmetric or antisymmetric channels.

At large $N$, the operators acquiring anomalous dimensions are of the schematic form,
\begin{equation}
\mathcal{O}_{pq}^\pm= \mathbb{P}_{I_1 I_2}^{\pm} \mathcal{O}_p^{I_1} \partial^l \square^{\frac{1}{2}(\tau-p-q)} \mathcal{O}_q^{I_2}
\end{equation}
where $\mathbb{P}_{ij}$ is an appropriate projector that projects onto symmetric or antisymmetric representations of the gauge group exchanged in the OPE.
For any given quantum numbers $\vec{\tau}=(\tau,b,l,a)$, the number of operators exchanged in the OPE can be represented with the number of pairs  $(pq)$ filling a rectangle \cite{Aprile:2018efk},
\begin{align}\label{ir}
{R}_{\vec{\tau}}:=  \biggl\{(p,q):  &\begin{array}{l}
	p=i+|a|+1+r\\q=i+a+1+b-r\end{array},   \notag \\
&	\left. \text{ for } \begin{array}{l}
	i=1,\ldots,(t-1)\\ r=0,\ldots,(\mu -1)\end{array}
	\right\}\,.
\end{align}
The rectangle $ {R}_{\vec{\tau}}$ consists of $d=\mu(t-1)$ allowed lattice points where
\be\label{multiplicity}
t\equiv \frac{(\tau-b)}{2}- \frac{(a+|a|)}{2},\,\,\,
\mu \equiv   \left\{\begin{array}{ll}
\bigl\lfloor{\frac{b+a-|a|+2}2}\bigr\rfloor \quad &a+l \text{ even,}\\[.2cm]
\bigl\lfloor{\frac{b+a-|a|+1}2}\bigr\rfloor \quad &a+l \text{ odd.}
\end{array}\right.
\notag
\ee
The picture below shows an example with $\mu=4, t=9$.
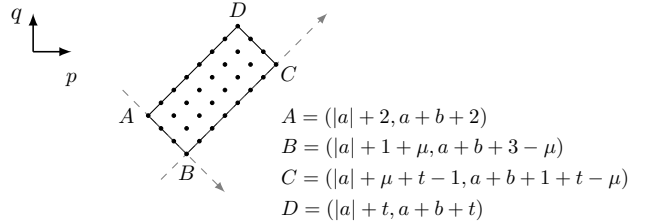
\be
\begin{tikzpicture}[scale=.34]
%
%
\def\prop{.5}
\def\shifthor{\prop*2}
\def\ptuno{(\prop*2-\shifthor,\prop*8)}
\def\ptdue{(\prop*5-\shifthor,\prop*5)}
\def\pttree{(\prop*9-\shifthor,\prop*15)}
\def\ptquattro{(\prop*12-\shifthor,\prop*12)}
%
\draw[-latex, line width=.6pt]		(\prop*1   -\shifthor-4,         \prop*14          -0.5*\shifthor)    --  (\prop*1  -\shifthor-2.5  ,   \prop*14-      0.5*\shifthor) ;
\node[scale=.8] (oxxy) at 			(\prop*1   -\shifthor-2.5,  \prop*16.5     -0.5*\shifthor)  {};
\node[scale=.8] [below of=oxxy] {$p$};
%
\draw[-latex, line width=.6pt] 		(\prop*1   -\shifthor-4,     \prop*14       -0.5*\shifthor)     --  (\prop*1   -\shifthor-4,        \prop*17-      0.5*\shifthor);
\node[scale=.8] (oxyy) at 			(\prop*1   -\shifthor-2,   \prop*16.8   -0.5*\shifthor) {};
\node[scale=.9] [left of= oxyy] {$q$};
%
\draw[] 								\ptuno -- \ptdue;
\draw[black]							\ptuno --\pttree;
\draw[black]							\ptdue --\ptquattro;
\draw[]								\pttree--\ptquattro;
\draw[-latex,gray, dashed]					(\prop*0-\shifthor,\prop*10) --(\prop*8-\shifthor,\prop*2);
\draw[-latex,gray, dashed]					(\prop*3-\shifthor,\prop*3) --(\prop*16-\shifthor,\prop*16);
%
%
\foreach \indeyc in {0,1,2,3}
\foreach \indexc  in {2,...,9}
\filldraw   					 (\prop*\indexc+\prop*\indeyc-\shifthor, \prop*6+\prop*\indexc-\prop*\indeyc)   	circle (.07);
%
%
\node[scale=.8] (puntouno) at (\prop*5-\shifthor,\prop*8) {};
\node[scale=.8]  [left of=puntouno] {$A$};   
\node[scale=.8] (puntodue) at (\prop*5-\shifthor,\prop*6+1.1) {};
\node[scale=.8] [below of=puntodue]  {$B$}; 
\node[scale=.8] (puntoquattro) at (\prop*13-\shifthor,\prop*16) {};
\node[scale=.8] [below of=puntoquattro] {$C$};
\node[scale=.8] (puntotre) at (\prop*9-\shifthor,\prop*11+.3) {};
\node[scale=.8] [above of=puntotre] {$D$}; 
%
%
\node[scale=.78] (legend) at (12,2) {$\begin{array}{l}  
													\displaystyle A=(|a|+2,a+b+2) \\[.1cm]
													\displaystyle B=(|a|+1+\mu,a+b+3-\mu) \\[.1cm]
													\displaystyle C=(|a|+\mu+t-1,a+b+1+t-\mu) \\[.1cm]
													\displaystyle D=(|a|+t,a+b+t) \\[.1 cm] \end{array}$  };
\end{tikzpicture}
\notag
\ee
This representation turns out to be particularly useful when we take into account $1/N$ corrections. In fact, operators on the same vertical line will continue to be degenerate at this order.
To see this, let us consider the OPE at genus zero. This is best cast in a matrix form \cite{Aprile:2017xsp}. First, arrange a $d \times d$ matrix of correlators
\begin{equation}
 \delta_{p_1 p_3}\delta_{p_2 p_4}G_{\text{disc},\vec{p}}^{\pm} \big |_{\text{long}} + \frac{1}{N}\mathcal{P} (x-y)(\bar{x}-y) \mathcal{A}_{\vec{p}}^\pm
\end{equation}
with the pairs $(p_1,p_2)$ and $(p_3,p_4)$ running over the same $R_{\vec{\tau}}$. Here, we denote by $\mathcal{A}_{\vec{p}}^\pm$ the inverse Mellin transform of the following Mellin amplitudes,
\begin{align}
\mathcal{M}_{\vec{p}}^{\pm}& =  \frac{1}{2}\bigl( \mathcal{M}_{\vec{p}}(1,2,3,4)\pm \mathcal{M}_{\vec{p}}(1,3,4,2) \bigr) \notag \\
& = \frac{1}{2}\frac{1}{{\bf s}+1} \left( \frac{1}{{\bf t}+1} \pm \frac{1}{{\bf u}+1} \right) .
\end{align}
The OPE equations then read
\begin{align} \label{OPEequations}
& \mathbf{C}_{\vec{\tau}}^\pm {\mathbf{C}_{\vec{\tau}}^\pm} ^T = \mathbf{L}_{\vec{\tau}}^{\pm}\,, \notag \\
& \mathbf{C}_{\vec{\tau}}^\pm {\bm{\eta}}_{\vec{\tau}}^\pm {\mathbf{C}_{\vec{\tau}}^\pm} ^T  = \mathbf{M}_{\vec{\tau}}^\pm\,.
\end{align}
Here $ \mathbf{L}_{\vec{\tau}}^\pm $ is a (diagonal) matrix of CPW coefficients of disconnected free theory defined by \eqref{discoeq}, while $\mathbf{M}_{\vec{\tau}}^\pm$ is a matrix of CPW coefficients of the $\log U$ discontinuity of $\mathcal{A}_{\vec{p}}^\pm$
\begin{equation}
\mathcal{P} (x-y)(\bar{x}-y)\mathcal{A}_{\vec{p}}^\pm \big |_{\text{log $U$}}= \sum_{\vec{\tau}}  M_{\vec{\tau}}^\pm \mathbb{L}_{\vec{\tau}}.
\end{equation}
Finally, ${\bm{\eta}}_{\vec{\tau}}^\pm$ is a diagonal matrix of anomalous dimensions and $\mathbf{C}_{\vec{\tau}}^\pm=\langle \mathcal{O}_p \mathcal{O}_q \mathcal{K}_{rs}^\pm \rangle$ is a matrix of three-point functions with two half-BPS and one double-trace operator. Here, we denote with $\mathcal{K}_{rs}^\pm$ the true two-particle operator in interacting theory, that differs by $\mathcal{O}_{pq}^\pm$, precisely because there is mixing. Note that, since $\mathcal{A}_{\vec{p}}^{\pm}$ can be written as a function of $\tilde{U}$ and $\tilde{V}$, the $SU(2)_L \times SU(2)_R$ representations contributing to $\mathbf{M}_{\vec{\tau}}^\pm$ can be reorganised into $SO(4)$ representations, while this is not so for the disconnected contribution $ \mathbf{L}_{\vec{\tau}}^{\pm}$.

It is simple to show, with some linear algebra, that the anomalous dimensions are the eigenvalues of the matrix $\mathbf{M}_{\vec{\tau}} \left( \mathbf{L}_{\vec{\tau}}^{\pm}\right)^{-1}$. 
By computing them for various quantum numbers, we find that the anomalous dimensions follow a very simple pattern,
\begin{equation}\label{YMandimensions}
\eta_{\vec{\tau}}^{\pm}= - \frac{2}{N} \frac{\delta_{h,j}^{(2)}\delta_{\bar{h},j}^{(2)}}{(l_{8d}^{\pm}+1)_4}
\end{equation}
where $l_{8d}$ is
\begin{equation}
l_{8d}^{\pm}= l+2(p-2)+ \frac{1\mp(-1)^{a+l}}{2}-|a|\,,
\end{equation}
and can be interpreted as a sort of effective $8d$ spin, the definition being dictated by the partial wave decomposition of the flat amplitude in $8d$ \cite{Caron-Huot:2018kta}.

Note that \eqref{YMandimensions} only depends on $p$, not $q$, or in other words, operators on the same vertical line in the rectangle will acquire the same anomalous dimensions.
We stress again that these are the anomalous dimensions associated to the double-trace operators exchanged in the amplitudes $\mathcal{M}_{\vec{p}}^{\pm}$: the gauge group enters the anomalous dimensions only through an overall constant which does not play any significant role in the computation.

Finally, note that in $AdS_5 \times S^5$ the anomalous dimensions read \cite{Aprile:2018efk}
\begin{equation}
\eta_{\vec{\tau}}= - \frac{2}{N^2} \frac{\delta_{h,\bar{h},j,\bar{j}}^{(4)}\delta_{h,\bar{h},\bar{j},j}^{(4)}}{(l_{10d}+1)_6}\,,
\end{equation}
where we recall that
\begin{equation}
\delta_{h,\bar{h},j,\bar{j}}^{(4)}\delta_{h,\bar{h},\bar{j},j}^{(4)} \equiv {\delta_{h,j}^{(2)}\delta_{\bar{h},\bar{j}}^{(2)}}{\delta_{h,\bar{j}}^{(2)}\delta_{\bar{h},j}^{(2)}}.
\end{equation}
Note that the numerator is doubled with respect to the $AdS_5 \times S^3$ case, as a consequence of the fact that supersymmetry is also doubled. Finally, let us also point out that the object $\delta_{h,\bar{j}}^{(2)}$ appearing ubiquitously is, perhaps with no much surprise, nothing but the anomalous dimension of the two-derivative sector in $AdS_2 \times S^2$ \cite{Abl:2021mxo}.

We conclude the section by commenting on the flavour structure of the correlator.
One way to deal with it is to decompose $t,u$ channel flavour structures (of both disconnected and tree-level correlators) in a basis of representations appearing in the tensor product of two adjoint representations in the $s$ channel. We then read off the coefficients associated to each flavour structure which are of the form 
\begin{align*}
& G^{I_1 I_2 I_3 I_4}_a \propto G^{I_1 I_2 I_3 I_4}_t+ G^{I_1 I_2 I_3 I_4}_u \qquad a \in \text{symm} \notag \\
& G^{I_1 I_2 I_3 I_4}_a \propto G^{I_1 I_2 I_3 I_4}_t- G^{I_1 I_2 I_3 I_4}_u \qquad a \in \text{anti}
\end{align*}
where $a$ runs over all symmetric (antisymmetric) representations in $\texttt{adj} \otimes \texttt{adj}$ with the proportionality coefficient depending on the specific group as well as the exchanged representation. Examples of such coefficients are given in \cite{Alday:2021ajh}.
The unmixing procedure can then be consistently carried for each $a$ separately. For the symmetric (antisymmetric) representations the relevant double-trace operators exchanged are of the type $\mathcal{O}_{pq}^+$ ($\mathcal{O}_{pq}^-$) with the respective anomalous dimensions proportional to $\eta_{\vec{\tau}}^{+}$ ($\eta_{\vec{\tau}}^-$). Actually, the only antisymmetric representation exchanged is the adjoint itself.

\subsection{Outlook and conclusions}
In the first part of this letter we have discussed colour-kinematics and BCJ relations between colour-ordered amplitudes of super gluons in $AdS_5 \times S^3$, by making use of the large $p$ formalism \cite{Aprile:2020luw}. We believe this formalism makes clearer the direct parallel with the flat space versions of these relations and that they hold for \emph{all} Kaluza-Klein modes. This, in turn, shows that, like in flat space, there is a precise relation between colour-kinematic duality and BCJ relations.

In the second part of the paper we have computed the anomalous dimensions in the large $N$ limit. As a consequence of the $8d$ hidden conformal symmetry, and in common with the analogous problems in $AdS_5 \times S^5$ and $AdS_3 \times S^3$, the anomalous dimensions turn out to have a residual degeneracy which is nicely captured by the vertical columns of the rectangular lattice $R_{\vec{\tau}}$ described in (\ref{ir}) and below. We were also able to give a number of formulae which interpolate results on the spectrum between different backgrounds of different dimensions.

These results open a lot of exciting possibilities. Firstly, as we mentioned  already in the introduction, the knowledge of the anomalous dimensions can be of use in bootstrapping loop corrections, beyond the lowest charge correlator studied in \cite{Alday:2021ajh}, and help in further exploring whether some features of the double-copy relations persist beyond tree level.
Moreover, following the procedure described in \cite{Abl:2020dbx}, one can imagine treating the theory of gluons as an effective model and introduce higher order $D^nF^4$ interactions, analogous to the higher curvature corrections present for gravitons in e.g. $AdS_5 \times S^5$. Much like the curvature corrections responsible for completing the Virasoro-Shapiro amplitude in $AdS_5 \times S^5$ \cite{Drummond:2020dwr,Aprile:2020mus}, such terms will induce a splitting of the residual degeneracy in the anomalous dimensions. Finally, the computation of open and closed string amplitudes in $AdS$ might give a clue on how KLT and world-sheet monodromy relations work in a curved spacetime.

\subsection{Acknowledgements}
We thank F.~Aprile, P.~Heslop, K.~Rigatos and X.~Zhou for providing important feedback and comments on the manuscript.
MS thanks D.~Bufalini, H.~Paul and S.~Rawash for useful discussions. 
JD is supported in part by the ERC Consolidator grant 648630 IQFT. RG is supported by an STFC studentship.
MS is supported by a Mayflower studentship from the University of Southampton.
\appendix

\subsection{Appendix: superconformal blocks}\label{longblocks}
We quickly review here the superconformal block technology needed in this letter. The long superconformal blocks \cite{Doobary:2015gia}, which capture the the non-protected multiplets exchanged in the OPE, are the simplest. They are the product of ordinary conformal and internal blocks for both $SU(2)$ factors. In our notation they take the form
\begin{equation}
\mathbb{L}_{\vec{\tau}}=\mathcal{P} (x-y)(\bar{x}-y)\biggl( \frac{\tilde{U}}{U} \biggr)^{p_3} \mathcal{G}_{\tau,l}(x,\bar{x}) \mathcal{H}_{b,a}(y,\bar{y})\,,
\end{equation}
where
\begin{align}
&   \mathcal{G}_{\tau,l}(x, \bar{x})= \notag \\
& \frac{(-1)^l}{(x-\bar{x})U^{\tfrac{p_{43}}{2}}} \Bigl( \mathcal{F}_{\tfrac{\tau}{2}+1+l}^+(x)\mathcal{F}_{\tfrac{\tau}{2}}^+(\bar{x})-\mathcal{F}_{\tfrac{\tau}{2}}^+(x)\mathcal{F}_{\tfrac{\tau}{2}+1+l}^+(\bar{x}) \Bigr),
\notag \\
&\mathcal{H}_{b,a}(y,\bar{y})= \frac{1}{\tilde{U}^{2-\tfrac{p_{43}}{2}}} \mathcal{F}_{-\tfrac{b}{2}-a}^-(y)\mathcal{F}_{-\tfrac{b}{2}}^-(\bar{y}),
\end{align}
with
\begin{equation}
\mathcal{F}_{h}^{\pm}(x)= x^h \!\!~_2F_1 \left[h \mp \tfrac{p_{12}}{2},h \mp \tfrac{p_{43}}{2}, 2 h \right](x).
\end{equation}
Here, $\mathcal{G}_{\tau,l}(x,\bar{x})$ are the standard 4d conformal blocks (up to a shift by 2 in the twist $\tau$) and $\mathcal{H}_{b,a}(y,\bar{y})$ are the internal blocks. The latter are the product of two $SU(2)$ spherical harmonics, one corresponding to the R-symmetry group $SU(2)_R$ and the other corresponding to the flavour group $SU(2)_L$. Finally, $\tau,l$ are, respectively, twist and spin, and $b,a$ label the different representation of $SO(4)$ and can be viewed as the analogues of twist and spin on the sphere.

Notice that the internal blocks are not invariant under $y \leftrightarrow \bar{y}$ exchange. In fact, the two $SU(2)$ have a different nature; in paricular, notice that free theory is also not invariant under $y \leftrightarrow \bar{y}$. This means that, unlike other theories like $AdS_5 \times S^5$ and $AdS_5 \times S^3$, the decomposition is extended to spherical harmonics with label $a<0$. In particular, for given charges $p_i$, we decompose a function in spherical harmonics labelled by two quantum numbers that we denote by $[ab]$. The values of $a$ run over the following set:
\begin{equation}
 -\kappa_{\vec{p}} \leq a \leq \kappa_{\vec{p}} \notag \\
\end{equation}
where
\begin{equation}
\kappa_{\vec{p}}=\frac{\min(p_1+p_2,p_3+p_4)-p_{43}-4}{2}
\end{equation}
is the `degree of extremality' and $p_{43}= p_4-p_3$.
For each value of $a$, $b$ runs over the set
\begin{equation}
  - \min(a,0) \leq \frac{b-p_{43}}{2} \leq (\kappa_{\vec{p}}-a+ \min(a,0) ).
\end{equation}

\bibliographystyle{apsrev4-1}

\end{document}